\pdfoutput=1
\documentclass[journal,10pt]{IEEEtran}  
\IEEEoverridecommandlockouts
\usepackage{cite}
\usepackage{indentfirst}
\usepackage{graphicx}
\usepackage{changepage}
\usepackage{stfloats}
\usepackage{amsfonts,amssymb}
\usepackage{bm}
\usepackage{algorithm}
\usepackage{algorithmic}
\usepackage{amsmath}
\usepackage{amsmath,amssymb,amsfonts}
\usepackage{times}
\usepackage{url}
\usepackage{cite}
\usepackage{textcomp}
\usepackage{xcolor}
\usepackage{color}
\usepackage{setspace}
\usepackage{enumitem}
\usepackage{float}
\usepackage{diagbox}
\usepackage[T1]{fontenc}
\usepackage[utf8]{inputenc}
\usepackage{authblk}
\usepackage{threeparttable}
\usepackage{booktabs}
\usepackage{footnote}
\usepackage{graphicx}
\usepackage{pifont}
\usepackage{subfigure}
\usepackage{mathrsfs}
\usepackage{makecell}
\usepackage{array}
\usepackage{booktabs}
\usepackage{lipsum}
\usepackage{multirow}
\usepackage{adjustbox}

\begin{document}
\title{UAV-Enabled Integrated Sensing and Communication: Opportunities and Challenges}

\author{
	Kaitao Meng, \textit{Member, IEEE}, Qingqing Wu, \textit{Senior Member, IEEE}, Jie Xu, \textit{Senior Member, IEEE}, Wen Chen, \textit{Senior Member, IEEE}, Zhiyong Feng, \textit{Senior Member, IEEE}, Robert Schober, \textit{Fellow, IEEE}, and A. Lee Swindlehurst, \textit{Fellow, IEEE}
	\thanks{{K. Meng is with the State Key Laboratory of Internet of Things for Smart City, University of Macau, Macau, 999078, China. }{(email: kaitaomeng@um.edu.mo).} Q. Wu and W. Chen are with the Department of Electronic Engineering, Shanghai Jiao Tong University, Shanghai 201210, China (email: \{qingqingwu, wenchen\}@sjtu.edu.cn). J. Xu is with the School of Science and Engineering, the Future Network of Intelligence Institute (FNii), and the Guangdong Provincial Key Laboratory of Future Networks of Intelligence, The Chinese University of Hong Kong (Shenzhen), Shenzhen, 518172, China, (email: xujie@cuhk.edu.cn). Z. Feng is with School of Information and Communication Engineering, Beijing University of Posts and Telecommunications, Beijing 100876, China (email: fengzy@bupt.edu.cn). Robert Schober is with the Institute for Digital Communications, Friedrich-Alexander University Erlangen-Nürnberg (FAU), 91054 Erlangen, Germany (email: robert.schober@fau.de).  A. L. Swindlehurst is with the Center for Pervasive Communications and Computing, University of California, Irvine, CA 92697, USA (email: swindle@uci.edu)}
	\thanks{Jie Xu's work was supported in part by the National Natural Science Foundation of China under grants No. U2001208, 92267202, the Shenzhen Fundamental Research Program under grant No. JCYJ20210324133405015. Wen Chen's work was supported by National key project 2020YFB1807700. Robert Schober's work was supported in part by the Deutsche Forschungsgemeinschaft (DFG, German Research Foundation) GRK 2680 – Project-ID 437847244. A. Lee Swindlehurst' work was supported by the U.S. National Science Foundation by grant CCF-2225575.
	}
}

\maketitle

%%%%%%%%%%%%%%%%%%%%%%%%%%%%%%%%%%%%%%%%%%%%%%%%%%%%%%%%%%%%%%%%%%%%%%%%%%%%%%%%
\begin{abstract}
Unmanned aerial vehicle (UAV)-enabled integrated sensing and communication (ISAC) has attracted growing research interests in the context of sixth-generation (6G) wireless networks, in which UAVs will be exploited as aerial wireless platforms to provide better coverage and enhanced sensing and communication (S\&C) services. However, due to the size, weight, and power (SWAP) constraints of UAVs, their controllable mobility, and the line-of-sight (LoS) air-ground channels, UAV-enabled ISAC introduces new opportunities and challenges. This article provides an overview of UAV-enabled ISAC and proposes various solutions for optimizing the S\&C performance. In particular, we first introduce UAV-enabled joint S\&C and discuss UAV motion control, wireless resource allocation, and interference management for ISAC systems employing single and multiple UAVs. Then, we present two application scenarios for exploiting the synergy between S\&C, namely sensing-assisted UAV communication and communication-assisted UAV sensing. Finally, we highlight several interesting research directions to guide and motivate future work.
\end{abstract}

%\begin{IEEEkeywords}
%Integrated sensing and communication (ISAC), unmanned aerial vehicle (UAV), fundamental trade-off, joint ISAC and trajectory design, mutual assistance.
%\end{IEEEkeywords}
%%%%%%%%%%%%%%%%%%%%%%%%%%%%%%%%%%%%%%%%%%%%%%%%%%%%%%%%%%%%%%%%%%%%%%%%%%%%%%%

\section{Introduction}
\par
Integrated sensing and communication (ISAC) has recently emerged as a candidate technology for sixth-generation (6G) wireless networks, in which wireless infrastructure and spectrum resources are shared to provide both sensing and communication (S\&C) services. By leveraging advanced multiple-input and multiple-output (MIMO) and millimeter-wave (mmWave)/terahertz (THz) technology, ISAC is expected to provide high-throughput, ultra-reliable, and low-latency wireless communications, as well as ultra-accurate and high-resolution wireless sensing for 6G \cite{Zhang2021Overview, Liu2020Joint}. This thus offers new opportunities for realizing environment- and location-aware applications for smart cities, smart manufacturing, autonomous driving, etc. However, conventional terrestrial ISAC networks can only provide sensing services within a fixed and limited range, as surrounding obstacles may block the line-of-sight (LoS) links to long-range targets, which leads to a seriously degraded sensing performance \cite{Zhang2021Enabling, liu2022integrated}.

\begin{figure*}[t]
	\centering
	\setlength{\abovecaptionskip}{0.cm}
	\includegraphics[width=17.8cm]{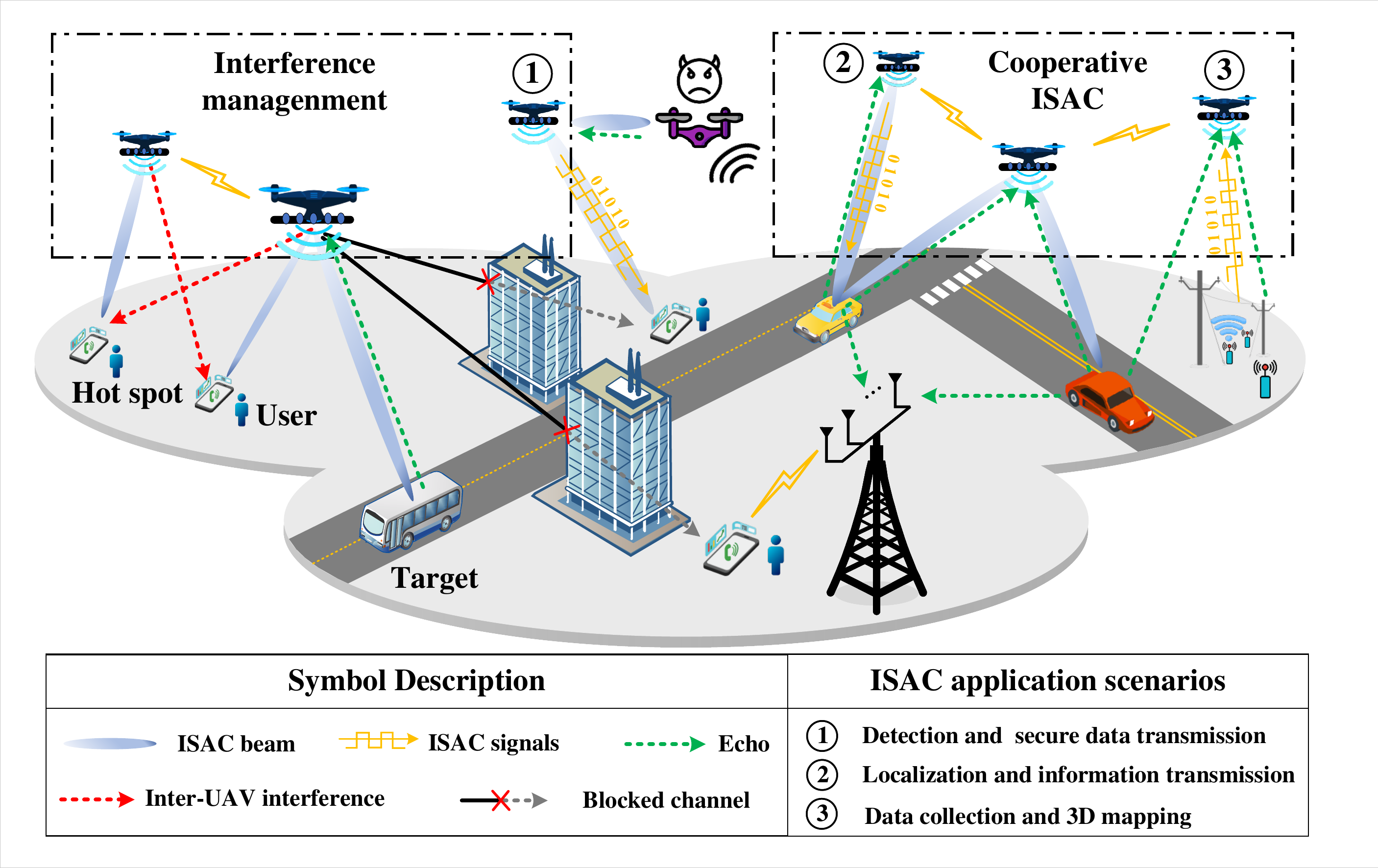}
	\caption{Application scenarios for UAV-enabled ISAC.}
	\label{figure1}
\end{figure*}

Motivated by the success of pilot projects on unmanned aerial vehicle (UAV)-enabled communications, such as AT\&T's flying COW and Nokia's F-cell \cite{Zeng2019Accessing}, there is a growing interest in employing UAVs as cost-effective aerial platforms to provide enhanced ISAC services supporting traffic accident rescue, non-authorized eavesdropper monitoring, and service enhancement in temporary hot spot areas, as illustrated in Fig.~\ref{figure1}. By exploiting the high mobility of UAVs in three-dimensional (3D) space and their strong air-ground LoS channels, UAV-enabled ISAC is expected to provide better S\&C coverage, more flexible surveillance, and enhanced S\&C performance compared to terrestrial ISAC. However, such a new aerial ISAC paradigm also introduces new design challenges. First, both fixed-wing and rotary-wing UAVs have to meet stringent constraints regarding their size, weight, and power, which limits their communication, sensing, and endurance capabilities. Second, strong air-ground LoS links inevitably incur severe interference in ISAC networks \cite{wang2020constrained}, which however can be exploited to extract rich target information such as location, velocity, and direction. Third, the flexible UAV placement/trajectory introduces new degrees-of-freedom (DoFs) for optimization, which makes the system design more complicated. Last but not least, unlike conventional UAV-enabled communications focusing on rate maximization, UAV-enabled ISAC systems need to incorporate (radar) sensing performance metrics (e.g., detection probability and estimation/recognition accuracy), sensing signal processing (e.g., echo signal processing and clutter interference suppression) \cite{Liu2022Survey}, and efficient cooperative mechanisms. As such, how to design UAV-enabled ISAC to achieve high S\&C performance and effective coordination among the UAVs is a new and challenging problem to address.

Given the above considerations, there is an urgent need to investigate joint S\&C design for UAV-enabled ISAC systems for improving spectrum efficiency, enabling hardware reuse, and reducing power consumption. Specifically, proper trajectory planning and resource allocation are needed to meet the distinct S\&C performance requirements and balance the performance-cost trade-off. For example, communication services are usually continuously provided for a period of time determined by the data volume, while sensing tasks tend to be performed with a certain sensing frequency depending on the targets' position and velocity, and the task's timeliness requirement. On the other hand, enforcing continuous sensing along with communication at all times may inevitably lead to higher energy consumption, a waste of spectrum resources, and stronger interference to communication users \cite{meng2022throughput}. Moreover, the use of multiple UAVs to collaboratively provide ISAC services is an efficient solution to further enhance the S\&C coverage and increase the integration gain, but such systems demand more sophisticated interference management \cite{wang2020constrained}.

Besides the integration gain obtained by the joint design of S\&C, mutual assistance of S\&C offers the potential to achieve a coordination gain in UAV-enabled ISAC, which enables sensing-assisted UAV communication and communication-assisted UAV sensing. For example, UAVs equipped with (radar) sensing capabilities can design their real-time trajectories and allocate communication resources based on their sensing results while incurring only a small signaling overhead. In turn, wireless communication provides an efficient means for UAVs to enhance their sensing data processing capabilities via, e.g., sensory data offloading and over-the-air computation \cite{Zhu2021OvertheAir}. 

In view of the above discussion, this article aims to provide a state-of-the-art overview of UAV-enabled ISAC, by identifying the related key challenges, discussing potential solutions, and presenting interesting directions for future research. To this end, Section \ref{SectionJointSandC} proposes new ISAC protocols and discusses UAV-enabled joint S\&C design for single- and multi-UAV systems, respectively. Sections \ref{SectionSensingAssistedC} and \ref{SectionCommunicationAssistedS} present novel concepts for sensing-assisted UAV communication and communication-assisted UAV sensing, respectively. Section \ref{Entensions} provides promising research directions for the integration of ISAC and UAVs. Finally, Section \ref{Conclusions} concludes the article.

\section{UAV-enabled Joint Sensing and Communication}
\label{SectionJointSandC}
This section discusses UAV-enabled joint S\&C, where UAVs serve ground communication users while concurrently detecting or estimating ground targets in relevant sensing areas. We distinguish between single- and multi-UAV systems.

\subsection{Single-UAV-Enabled ISAC} 

While sensing and communication functionalities could be time multiplexed, improved performance is expected if both services can operate as needed, and possibly simultaneously. Therefore, in this subsection, we present new transmission protocols, novel resource allocation strategies, and UAV trajectory designs for UAV-enabled joint S\&C. 

\subsubsection{ISAC Frame Protocol Design}

\begin{figure}[h!]
	\centering
	\setlength{\abovecaptionskip}{0.cm}
	\includegraphics[width=8.2cm]{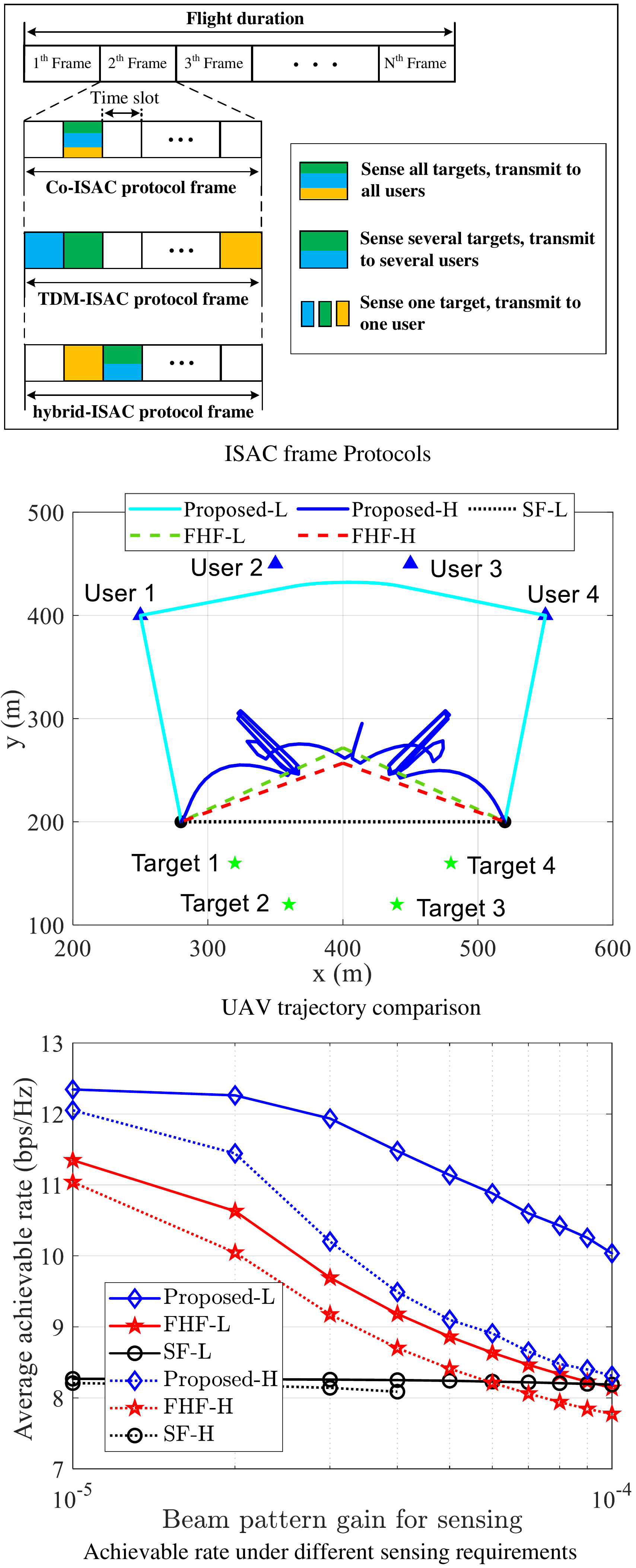}
	\caption{Illustration of ISAC protocols and comparisons of UAV trajectory and achievable rate.}
	\label{figure2}
\end{figure}

Suppose that unified ISAC waveforms or beams are employed to sense multiple targets, for which the received sensing signal-to-clutter-and-noise ratio (SCNR) \cite{liu2022integrated} or the sensing beampattern \cite{Liu2020JointTransmit} are adopted as performance metrics. Since communication is generally continuously required while sensing tasks are often performed periodically, specific ISAC frames should be designed to facilitate the resource allocation and trajectory optimization, as shown in the top subfigure of Fig.~\ref{figure2}. During each ISAC frame, different targets may be sensed simultaneously or separately at least once. Accordingly, the ISAC frame protocols can be classified into the following three categories. 
\begin{itemize}[leftmargin=*]
	\item \textit{Co-ISAC}: During each ISAC frame, all targets are sensed simultaneously at least once, and thus the ISAC beams need to be radiated divergently to cover all targets and users at the same time, as shown in the top subfigure of Fig.~\ref{figure2}. Due to the potentially stringent sensing requirements for all targets, the UAV trajectory design in this case is less flexible since the transmit power has to be divided into multiple directions for both S\&C.
	\item \textit{TDM-ISAC}: Multi-target sensing is performed in a time division multiplexing (TDM) manner along with communication functions, i.e., in each time instant unified waveforms/beams only cover one intended target (instead of all targets in \textit{Co-ISAC}) together with one communication user. In this case, echo signals from other targets become clutter/interference for the intended target sensing. On the other hand, a target and users with small angular separation tend to be jointly served to improve energy efficiency, since the leakage power of the sensing beam towards a user can be utilized for information transmission \cite{Liu2020Joint, Liu2020JointTransmit}.
	\item \textit{Hybrid-ISAC}: This protocol is a combination of \textit{Co-ISAC} and \textit{TDM-ISAC}. In this design, multiple targets are grouped based on their locations. Accordingly, \textit{Co-ISAC} is performed within each group to improve intra-group sensing efficiency while \textit{TDM-ISAC} is implemented among different groups to avoid inter-group interference. By properly optimizing the target grouping, this hybrid protocol is expected to outperform the \textit{Co-ISAC} and \textit{TDM-ISAC} protocols in terms of efficiency and cost. 
\end{itemize} 
The three above protocol designs have advantages and disadvantages, and their relative performance depends on various factors such as the S\&C quality-of-service (QoS) requirements, the locations of users/targets, and their mobility \cite{Keskin2021MIMO}. How to optimize the protocol design to enhance the S\&C performance needs further investigation.
\subsubsection{Joint Resource Allocation, Waveform, and Deployment/Trajectory Design}

Unlike conventional terrestrial ISAC systems, in UAV-enabled ISAC systems, optimal resource allocation and waveform design are deeply influenced by the UAV deployment/trajectory, since the angular separations between users/targets change with the UAV location. Therefore, to achieve high S\&C performance, user association, transmit beamforming, and the UAV trajectory must be jointly designed to maximize communication performance while ensuring the required sensing power and sensing frequency \cite{meng2022throughput}. Solutions to this problem can be generally divided into optimization-based and learning-based methods \cite{Zeng2019Accessing}. However, finding the optimal solution to the resulting joint optimization problem is challenging, since the beamforming design and UAV trajectory are closely coupled in multiple nested transcendental functions and integer optimization is required, e.g., for user association and target allocation \cite{meng2022throughput}. To tackle this issue, a two-layer penalty-based algorithm was proposed to decompose the involved coupled integer optimization variables for finding high-quality solutions \cite{meng2022throughput}. %in \cite{lyu2021joint}, a trust region-based algorithm to jointly optimize the beamforming and the UAV deployment/trajectory was developed by employing successive convex approximation (SCA) to transform the non-convex objective function and constraints into convex ones in each iteration. Furthermore, taking into account the frequency requirements of the sensing tasks, 

To demonstrate the effectiveness of the algorithm outlined above, the middle subfigure of Fig.~\ref{figure2} illustrates various UAV trajectory designs and the corresponding achievable communication rates under the TDM-ISAC protocol, for a scenario with 4 users and 4 targets. The parameters are set based on practical system requirements and related references \cite{Zeng2019Accessing}. In particular, the number of antennas at the UAV is 16, and the UAV's maximum horizontal flight speed is 30 m/s with a flight altitude of 40 m and a flight duration of 40 s. In addition, the channel gain at a reference distance of 1 m and the noise power at each user are set to $-30$ dB and $-70$ dBm, respectively. The maximum transmit power is 0.1 W, and the length of the time slots is 0.25 s. In the middle subfigure of Fig.~\ref{figure2}, two benchmarks are considered: 1) Straight flight (SF): The UAV flies from the initial location to the final location along a straight line at a constant speed of 6 m/s; 2) flight-hover-flight (FHF): The UAV flies directly at its maximum speed from the initial location to the optimal location, where the UAV can transmit with the maximum achievable rate, hovers at the optimal location for a certain period of time, and then flies straight to the final location. Our proposed scheme for a high (low) required sensing frequency is referred as to Proposed-L (Proposed-H), similar to the benchmark schemes. Here, the high (low) sensing frequency is 0.2 Hz (0.025 Hz). Specifically, the sensing frequency refers to the reciprocal of the interval between two sensing times. It is observed that as the sensing frequency increases, the UAV's trajectory gradually shrinks from a longer arc towards the users to several turn-back sub-trajectories between the targets and the users. The bottom subfigure of Fig.~\ref{figure2} unveils a fundamental trade-off between sensing frequency and communication rate in UAV-enabled ISAC systems.

\begin{table*}[t] 
	\centering
	\caption{Comparison between single-UAV-enabled ISAC and multi-UAV-enabled ISAC.}
	\label{Table1}
	\begin{tabular}{ |p l | l | l| l|}
		\hline
		\multicolumn{2}{|c|}{\bf{Classification}} & {\makecell[c]{\bf{Advantages}}} & {\makecell[c]{\bf{Weaknesses}}} \\ \cline{2-4}
		\hline
		\multicolumn{2}{|c|}{{Single-UAV-enabled ISAC (Section II-A)}} & {\makecell[c]{Low-cost, \\ less interference  }} & {\makecell[c]{Small coverage, \\ high latency  }} \\ \hline
		\multicolumn{1}{|c|}{\multirow{6}{*}{\begin{tabular}[c]{@{}c@{}}Multi-UAV-enabled ISAC\end{tabular}}} & {\makecell[c]{Coordinated interference  \\ management (Section II-B1)}}  & {\makecell[c]{Large coverage, \\ low complexity, \\ less overhead  }} & {\makecell[c]{Under-utilized echo, \\ strong interference \\ in open space }} \\ \cline{2-4}
		\multicolumn{1}{|c|}{}   & {\makecell[c]{Cooperative ISAC \\ (Section II-B2)}}  & {\makecell[c]{Multi-directional \\ observation, richer \\  target  information, \\ deep coordination, \\ more flexibility   }} & {\makecell[c]{Large overhead, \\ high complexity, \\ strict time
				\\	synchronization  }} \\ 
		\hline
	\end{tabular}
\end{table*}

The complexity of the above trajectory design methods may become intractable for long flying periods. In fact, how to design a low-complexity trajectory achieving satisfactory performance is an open problem of high practical interest. A possible solution for this challenge is to partition the whole period into a number of ISAC frames with limited duration. In this manner, for periodic sensing tasks, we can obtain the trajectory for one ISAC frame, based on which the trajectories for the other ISAC frames can be constructed, thereby reducing the algorithmic complexity \cite{meng2022throughput}.

\subsection{Multi-UAV-Enabled ISAC}

In single-UAV-enabled ISAC, the achievable S\&C performance may be low for geographically distributed and time-critical tasks, due to the limited sensing range and communication rate of a single UAV. This thus motivates the development of effective multi-UAV collaboration schemes to further improve resource efficiency. Compared to the single-UAV scenario, multi-UAV-enabled ISAC requires mitigation of the potentially severe inter-UAV interference caused by the strong LoS-dominant air-ground channels. To allow for different levels of cooperation among UAVs, we consider two cases, namely coordinated interference management and cooperative ISAC. Their respective advantages and weaknesses are compared in Table \ref{Table1}.

\subsubsection{Coordinated Interference Management} 
\label{CoordinatedInterference}
In this case, each UAV serves the users and targets assigned to it, and different UAVs serve different users and targets. The UAVs may cause strong interference to adjacent unassociated users/targets, thus limiting the S\&C range and performance \cite{Chen2020Performance}. It is therefore of paramount importance to develop advanced countermeasures for managing such interference. One viable solution is to exploit the mobility of the UAV together with beamforming design and power control for reducing inter-UAV interference. Intuitively, sufficiently separated users/targets (i.e., the users'/targets' angular separations exceed the angular resolution of the antenna array installed on one UAV), are preferably served simultaneously by different UAVs, especially in poor scattering environments. The main reasons for this are that the interference among UAVs caused by the side lobes of communication beams is greatly reduced due to the low correlation of the user channels and that the received signals reflected from more separated targets are distinguishable by one UAV. Furthermore, obstacles in the surrounding environment can even be utilized for interference reduction through proper deployment/trajectory design. As illustrated in Fig.~\ref{figure1}, each UAV tends to hover at an optimized location that has LoS links to its associated users/targets but blocked LoS links to unassociated users/targets, thus enhancing the S\&C performance while minimizing the interference. This thus leads to a multi-UAV collaboration gain.

\subsubsection{Cooperative ISAC}
In cooperative ISAC, multiple UAVs perform distributed radar sensing and coordinated wireless communications with a higher degree of collaboration, thus enabling the combination of distributed MIMO radar and aerial coordinated multi-point (CoMP) transmission/reception compared to coordinated interference management. In this case, UAVs are also allowed to act as dedicated transmitters/receivers and send/receive correlated signals for collaborative S\&C. From the sensing perspective, by sharing or fusing the sensing results of different UAVs, larger sensing coverage, more diverse observation angles, and more accurate target parameter estimates are obtained. In addition, the received signals of all UAVs can be collected and fused at a central UAV or at on-ground base stations (BSs), and then the sensing results can be fed back to the different UAVs. Note that, for cooperative ISAC, the overhead required for information exchange is larger than that for coordinated interference management described in Section \ref{CoordinatedInterference}. Furthermore, for cooperative ISAC, the geometric dilution of precision (GDOP), as an important factor of positional measurement precision \cite{wang2020constrained}, needs to be optimized to realize a large distributed MIMO gain. From the communication perspective, by exploiting the benefits of the adjustable distributed antenna array created by multi-UAV systems, high spectrum efficiency can be achieved with the help of CoMP. 

It is worth noting that NLoS links are exploitable for communication with the served users, whereas typically only LoS links are exploited for sensing and NLoS links are treated as unfavorable interference. Accordingly, UAVs at higher altitudes and in more open environments are more likely to have strong LoS links to the targets sensed by neighboring UAVs, and thus more reflected signals can be utilized for collaborative sensing; on the contrary, multi-user communications may suffer from more potentially harmful interference and channels with fewer DoFs due to LoS-dominated links. Therefore, UAV deployments with strong LoS links to the intended targets as well as a sufficiently large number of NLoS links to communication users to create high-rank MIMO channels are preferable, leading to a fundamental trade-off between S\&C performance. Furthermore, the ground BSs can assist in radar signal processing and interference cancellation for communication signals in multi-UAV-enabled ISAC networks, as shown in Fig.~\ref{figure1}. Nonetheless, such distributed multi-static ISAC systems pose several new challenges that need to be addressed, including the high signaling overhead and strict time synchronization requirements. Therefore, more in-depth studies are needed to unveil the most suitable approach for realizing efficient and distributed multi-UAV-enabled ISAC. 

\section{Sensing-assisted UAV Communication}
\label{SectionSensingAssistedC}
\begin{figure}[t] 
	\centering
	\setlength{\abovecaptionskip}{0.cm}
	\includegraphics[width=8.2cm]{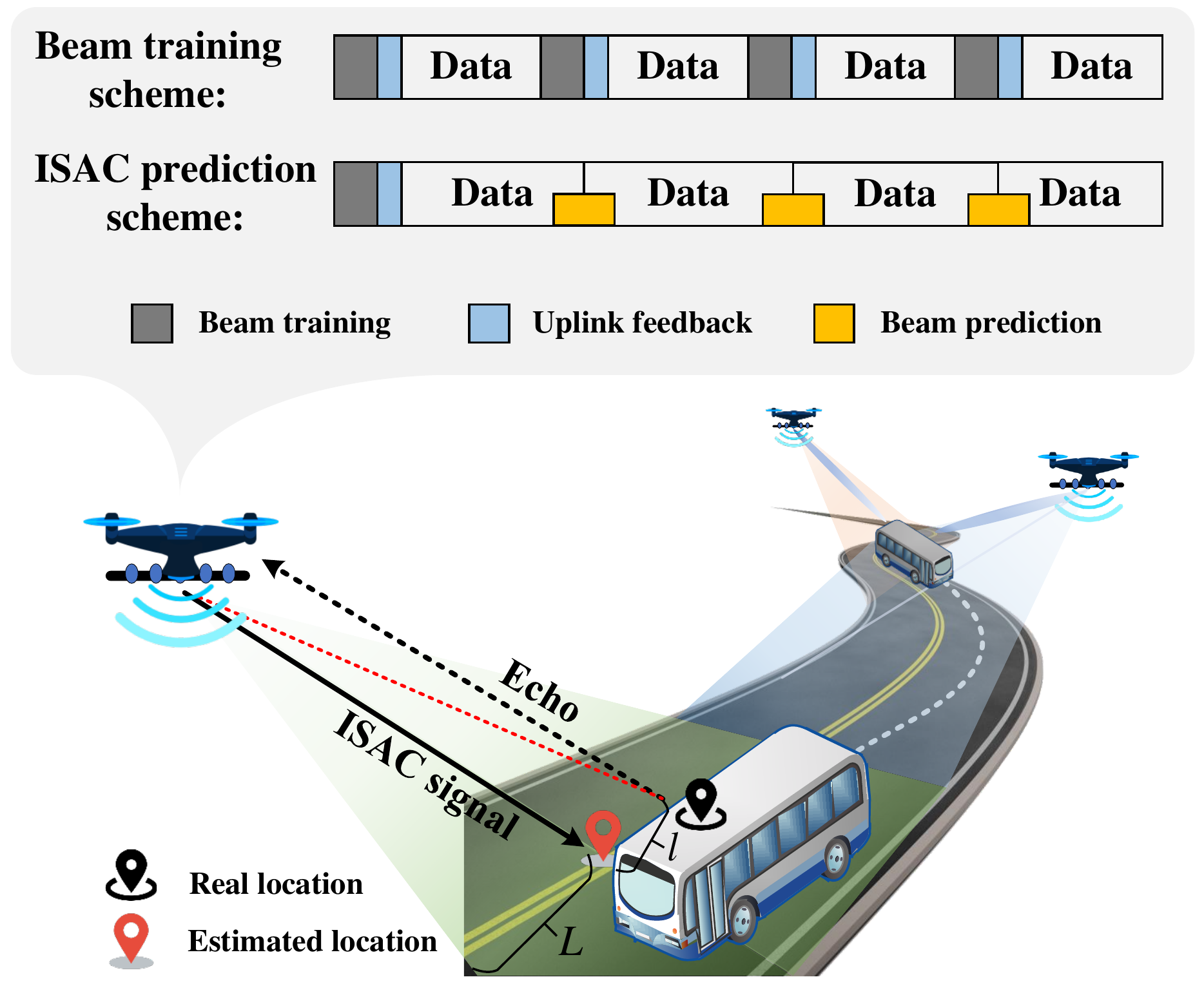}
	\caption{Sensing-assisted UAV communication.}
	\label{figure4}
\end{figure}

Sensing can provide the capability to see the physical world for future wireless networks, which in turn can potentially enhance their communication performance \cite{Liu2020RadarAssisted}. For instance, instead of relying on sending pilots to the receivers and feeding back channel estimates to the transmitter (or performing channel estimation at the BS based on the pilots sent by users), the signals reflected by the served ISAC users with sensing and communication requirements can be directly utilized for localization and/or channel estimation. This thus helps to reduce the signaling overhead and yields a performance improvement, which gives rise to a new type of sensing gain. However, it remains an open problem how to quantitatively measure such sensing gain and how to fully exploit it for maximization of the communication performance by optimizing the UAV trajectory and/or beamforming. To find answers to these questions, we consider the UAV-to-ground vehicle communication scenario shown in Fig. \ref{figure4}, where the communication performance improvement introduced by sensing is analyzed. 

\subsection{Sensing Gain} 
Instead of downlink pilots or uplink feedback, the served ground vehicle's information, e.g., location, velocity, and angle, can be extracted from the reflected ISAC signals for use in beam tracking and beam alignment. To shed some light on the communication performance improvement achieved via sensing, the rate gain realized by ISAC prediction over conventional beam training is analyzed as follows. First, for ISAC prediction, the estimated vehicle location error may lead to beam misalignment, and the corresponding impact on the received communication signal-to-noise ratio (SNR) decreases exponentially with respect to (w.r.t.) the ratio of the angle estimation error $l$ (rad) and the equivalent beamwidth $L$ (rad) (c.f. Section II-B in \cite{Chang2022Integrated}). For comparison, for conventional beam training (c.f. the top subfigure in Fig.\ref{figure4}), the achievable rate of the served user is the product of two terms: the time ratio $1 - \alpha$, which accounts for the overhead introduced by the downlink pilots, and the communication rate, which accounts for the SNR loss $\beta_t$ caused by beam misalignment. The communication performance improvement gained with sensing, namely the sensing gain, can thus by characterized as the difference between the achievable rate of the proposed ISAC prediction scheme and that of the conventional beam training scheme.  %Then, the corresponding SNR is the product of performance loss caused by beam misalignment and the maximum SNR with perfect beam alignment. 

\begin{figure}[t]
	\centering
	\setlength{\abovecaptionskip}{0.cm}
	\includegraphics[width=8.2cm]{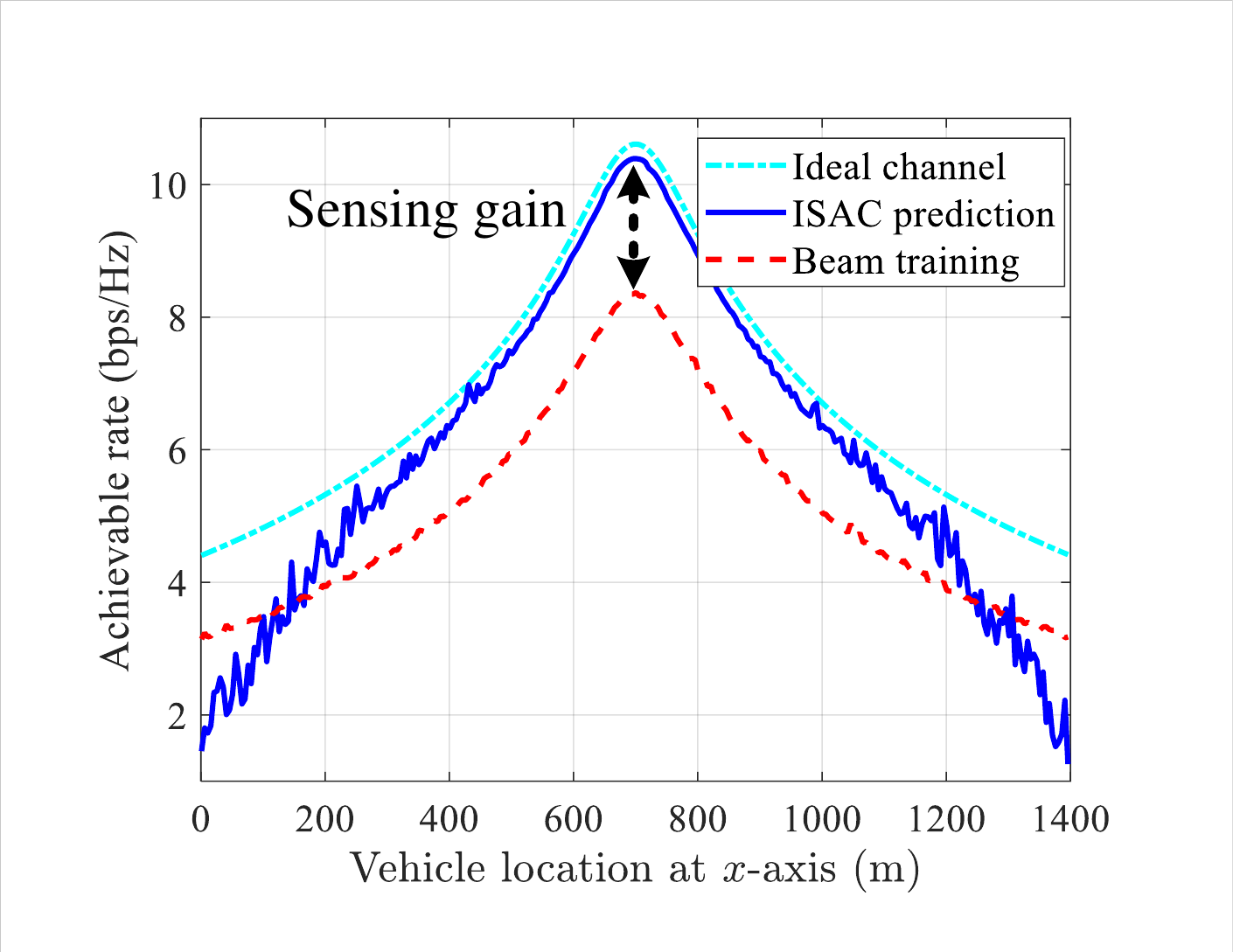}
	\caption{Sensing gain achieved by ISAC prediction over conventional beam training.}
	\label{figure5}
\end{figure}

Based on the above discussion, the more (less) accurate the target (channel) estimation, the larger the sensing gain that can be achieved. For LoS-dominated channels, the location estimation error $l$ is generally a function of the fourth power of the link distance between the UAV and the ground vehicle due to the round-trip path loss of the reflected signals, while the SNR loss of conventional beam training schemes depends on the received signal power at the ground vehicle. Thus, the rate gain realized by sensing-assisted communication is expected to decrease as the link distance increases. This is illustrated in Fig.~\ref{figure5}, where the end-to-end spectrum efficiency of conventional beam training and ISAC prediction are plotted for a setup, where the ground vehicle moves along the $x$-axis and the UAV is hovering at $x = 700$ m with a flight altitude of 80 m and a constant beam width. Fig.~\ref{figure5} reveals that a higher sensing gain is achieved when the ground vehicle is closer to the UAV, as the accuracy of ISAC prediction decreases and begins to fluctuate as the echo signal power becomes weaker. As a result, exploiting the UAV's mobility to shorten the link distance not only reduces the large-scale path loss but also strengthens the performance improvement gained from sensing. We note that joint beamwidth and UAV trajectory design is a promising approach to further improve ISAC performance.

For general multi-UAV scenarios, collaborative sensing potentially leads to significant communication performance improvement but requires sophisticated cooperation schemes. In particular, how to realize efficient and reliable sensing data exchange and fusion among multiple UAVs for high-quality and seamless communication coverage is an open problem that requires further investigation.
 
\subsection{Sensing-assisted Beam Tracking} 

How to achieve precise target tracking and a high beamforming gain for communication is also an open problem. Specifically, for long-distance users that can be modeled as point-like objects, the sensing/radar beam should be designed as narrow as possible to accurately point towards its receive antennas, thereby providing both high beamforming gain for communication and excellent angular resolution for sensing. On the other hand, nearby users are not points but rather angularly extended objects, and in this case, a wider beam is preferred to cover the extended object while a narrower beam towards the receiver antennas can realize potentially higher communication performance. One possible approach to achieve an efficient balance between S\&C is to employ a dynamic waveform to adjust the width and center of the ISAC beam in real time according to the relative position of the receive antennas w.r.t. the estimated contour of the object. For example, in the S\&C stage, the beamwidth may be designed to cover the entire object to guarantee sensing accuracy with relatively low communication performance \cite{Liu2020RadarAssisted}, while in the communication-only stage, a narrower beam can be adopted to align with the receive antennas based on the prior knowledge of the antennas' locations. In addition, for users with high mobility, it is advantageous for UAVs to use wide beams to provide reliable and effective target tracking at a possibly large distance to the target, while the communication performance can be improved by adopting narrow beams close to the target. This leads to a fundamental trade-off between communication throughput and sensing reliability for joint beamwidth and UAV trajectory design. Therefore, how to provide reliable beam tracking and enhanced communication performance by exploiting the mobility of UAVs and beamwidth design is a new and practically important problem.	

\subsection{Sensing-assisted Predictive Resource Allocation}

Although multi-UAV-enabled ISAC is promising for performance and coverage extension, it causes several practical challenges for resource allocation and user scheduling at the network level, such as dynamic load balancing and seamless coverage. For example, in multi-UAV networks, some UAVs may suffer from heavy S\&C traffic loads while others may have only light loads, due to the uneven distribution and mobility of the users. This thus seriously degrades the service time and quality due to the limited energy and resources of each UAV. One possible solution is to allow the UAVs to actively/passively monitor the served users' state (e.g., position and velocity) by analyzing their reflected signals, and then predict their trajectories based on the measured information. Then, these results can be further exploited to optimize the network resource allocation and user scheduling, thus achieving high-quality service by reserving resources and communication data for the users in advance. There are still many open and challenging issues for seamless coverage and connectivity in multi-UAV networks, especially in urban environments with many potential obstructions. Specifically, how to jointly design the dynamic UAV deployment and resource allocation to provide seamless service is a crucial challenge.

\section{Communication-assisted UAV Sensing}
\label{SectionCommunicationAssistedS} 
Besides sensing-assisted UAV communication, the communication functionality can also assist sensing to enhance the sensing robustness, efficiency, and accuracy.

\subsection{Data Offloading}

\begin{figure}[t]
	\centering
	\setlength{\abovecaptionskip}{0.cm}
	\includegraphics[width=8.2cm]{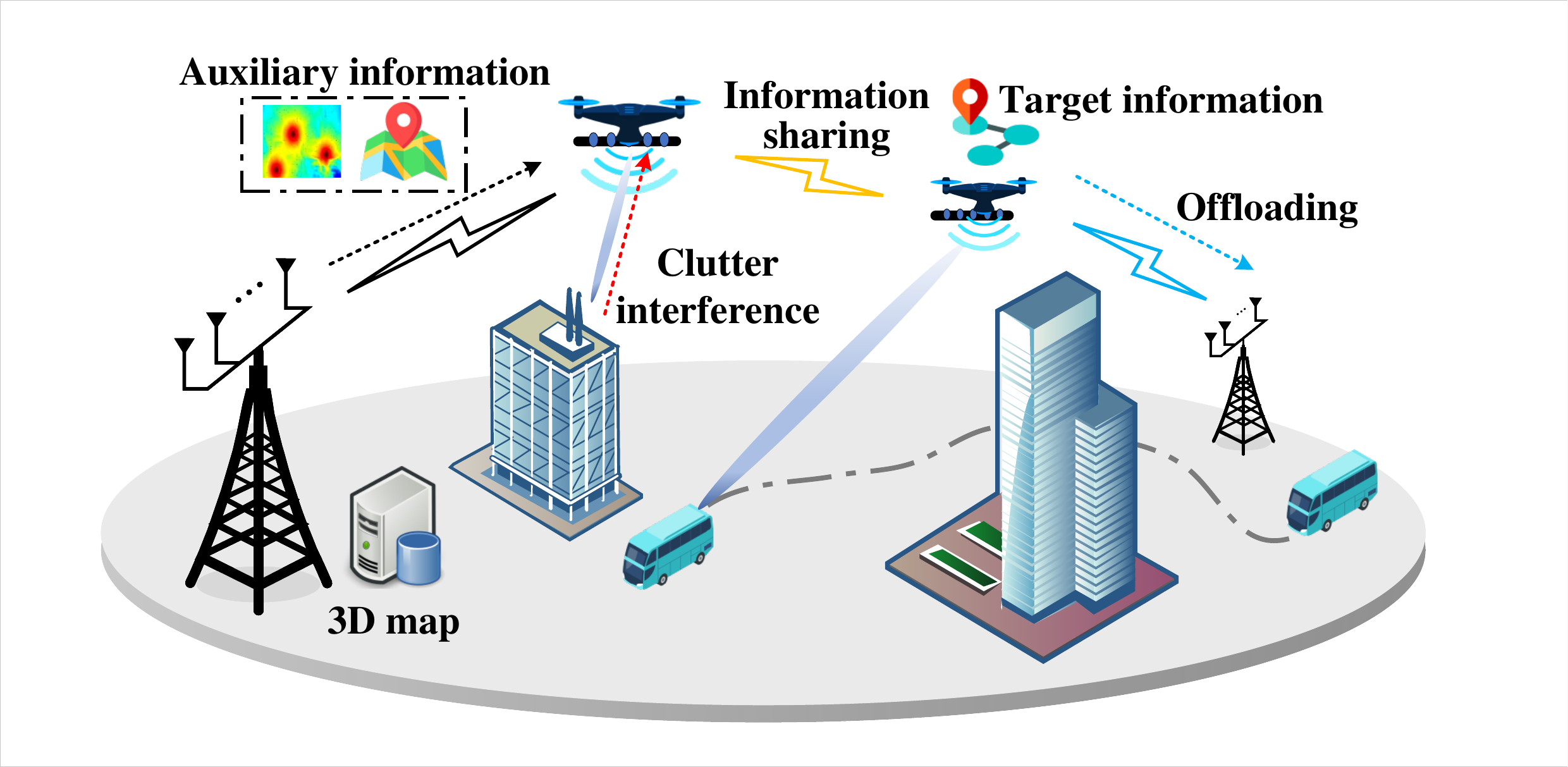}
	\caption{Communication-assisted UAV sensing.}
	\label{figure6}
\end{figure}

As the sensing results are generally needed for subsequent processes, two challenges for UAVs performing sensing tasks in practice are their limited computational capabilities and the low latency requirements for data processing. For example, processing all the received echoes locally at the UAV may be too time-consuming to meet the latency requirements of delay-sensitive ISAC missions, such as target tracking. To tackle this problem, one viable solution is to offload some computationally-intensive sensing tasks (e.g., in form of raw data or processed data) to nearby edge servers (e.g., at ground BSs or a central UAV with powerful computing capabilities), as shown in Fig.~\ref{figure6}. By judiciously selecting the computing nodes (e.g., those with strong LoS links to the UAV) and scheduling multi-dimensional resources (e.g., communication resource allocation and computation offloading optimization), ISAC services can be provided more efficiently ensuring better timeliness. However, how to balance the associated energy consumption and transmission/processing latency requires further study. Moreover, due to the potentially large amounts of sensory data and limited link capacity, advanced compression methods may be applied to pre-process the sensing results and reduce the transmission burden. Alternatively, multiple UAVs may form multi-hop links for collaboratively relaying and offloading sensing tasks.

\subsection{Information Sharing and Fusion}
Considering the limited sensing range and performance of a single UAV, another solution to improve the sensing performance is to allow multiple UAVs to share and integrate their information for joint processing. For example, individually estimated information regarding the users' positions and velocities may be shared among UAVs, and thus the sensing mission assignment can be made more efficient in the next ISAC frame. By sharing the users' direction of motion and changes in the surrounding environment, a multi-UAV system with maneuverability can collaboratively provide seamless coverage and tracking. In addition, through information sharing, the waste of resources caused by repetitive target detection and excessive target searching is avoided. Furthermore, a UAV or a ground BS can serve as a data center for the collection and fusion of sensing results, thus improving the sensing accuracy and obtaining richer target information. However, information sharing/exchange also introduces transmission latency and consumes communication resources. Hence, how to design a low-cost and highly-efficient data sharing/fusion strategy to improve the network sensing performance is an open and challenging issue. Since in practice the communication rates are often severely limited, it may be difficult to meet the stringent sensing latency requirements, especially for wireless data aggregation in swarm UAV scenarios. A promising approach for improving data fusion efficiency is to apply over-the-air computation \cite{Zhu2021OvertheAir}, which exploits the waveform superposition property of wireless channels to realize over-the-air aggregation of data simultaneously transmitted by multiple UAVs, without the need for separate data demodulation and fusion processes.
%Second, the handover criteria are more complex. Specifically, besides the signal strength which is normally the criterion for handover in communication, angles (AOA/AOD) and moving direction also matter a lot for sensing.
\subsection{3D Map Assistance}
One challenging issue related to UAV sensing arises from environmental obstacles, which could either block LoS sensing links or cause clutter interference. For example, when a UAV flies to an unknown area to perform an ISAC mission, the UAV-ground channels may be occasionally blocked by high-rise buildings in urban areas, which degrades the S\&C performance. To overcome this issue, one possible solution is to employ an environment map constructed based on historical measurements. For example, nearby BSs or edge servers may transmit a stored 3D map of the surrounding environment to the UAV, and based on this map, the states of the links between the UAV and the targets can be predicted. In turn, the map can be further updated based on the current sensing results. However, relying on map information only does not allow to address the dynamics of the environment. To tackle this problem, one viable method is to combine the offline LoS modeling and online sensing information (e.g., positions of the UAV and obstacles) to more accurately determine whether there exists an LoS link between the UAV and a given target location. This enables the UAV to design its real-time trajectory to ensure LoS links to the served users for providing enhanced and reliable ISAC services. Furthermore, it is also possible to extract auxiliary information for sensing based on a 3D environment map, such as the features of the explored/served areas and potential clutter. With such information, the UAV is able to create awareness of the environment around it and reduce/cancel clutter interference for facilitating target sensing, as illustrated in Fig.~\ref{figure6}.

\section{Directions For Future Research}
\label{Entensions}
Some open issues and challenges related to the integration of ISAC and UAVs are discussed in the following sections.

\subsection{ISAC for UAVs}
ISAC networks can also be utilized to monitor and manage network-connected UAVs, especially UAVs at low altitudes. For network-connected UAVs, ISAC signals emitted by ground BSs can be used for tracking the UAVs and thus enhancing the communication performance through efficient beam prediction. By exploiting the UAVs' reflected signals, a more reliable cellular connection can be realized by proper resource allocation and trajectory design. However, the strong UAV-ground LoS links inevitably increase the interference to terrestrial users/BSs \cite{liu2022integrated}. This motivates the development of new techniques for cooperative interference management and cancellation for heterogeneous ISAC networks.

\subsection{IRS-assisted UAV-enabled ISAC}
Intelligent reflecting surface (IRS) is a promising technology to reconfigure wireless channels by exploiting smart reflections by massive low-cost reflecting elements. By exploiting an IRS, a virtual LoS link between a UAV and blocked users can be established to enlarge the UAV's coverage area. This in turn provides higher flexibility for UAV deployment/trajectory design to achieve better S\&C performance. Thus, IRS and UAV can synergistically improve S\&C performance by proactively jointly altering the wireless communication channel via phase shift and trajectory design, respectively. However, the signaling overhead required for channel estimation is expected to be significant and the joint system design may entail high complexity.

\subsection{Secure UAV ISAC}

UAV-enabled ISAC systems increase the risk of eavesdropping and jamming attacks due to the LoS-dominated air-ground channels. In addition, unauthorized malicious UAVs pose a new security threat to ground ISAC networks. As such, how to effectively safeguard the legitimate S\&C users (e.g., preventing the target location and user information from being eavesdropped) and how to efficiently protect the S\&C services (e.g., accurate sensing and reliable communication) against malicious attacks are new and challenging problems to address. Combining information signals with artificial noise is a promising approach for target/eavesdropper tracking \cite{Lu2020UAVCellular}, but providing secure ISAC services is still challenging due to the difficulty in determining the locations and channels of the eavesdroppers.

\subsection{UAV ISAC Meets Artificial Intelligence}

Artificial intelligence (AI)-based designs are promising options for coping with such highly dynamic scenarios while avoiding the time-consuming iterations of traditional optimization algorithms \cite{Lu2020UAVCellular}. By integrating sensing information, such as the states of the environment, into the AI algorithm, future network states can be predicted, which allows UAVs to adjust their actions in an online manner. In turn, ISAC can provide training data for new AI-enabled applications via wireless network sensing. To properly train AI models using ISAC data from distributed UAVs while preserving their privacy, federated learning (FL) can be an efficient solution, where each participating UAV updates its local AI model based on its own local ISAC data, and then sends the updated parameters to a central server for updating the global AI model. However, how to efficiently integrate the training algorithm and the ISAC process is an interesting open problem. 

\section{Conclusions}
\label{Conclusions}
In this article, we have discussed UAV-enabled ISAC to realize an integration gain and facilitate mutual support between S\&C. New design considerations and key challenges have been highlighted for UAV-enabled ISAC networks. Coordinated interference management and cooperative ISAC have been proposed for performance improvement in multi-UAV-enabled ISAC networks. Two representative examples for ISAC coordination gains, i.e., sensing-assisted UAV communication and communication-assisted UAV sensing, have been presented to demonstrate the complementary nature of S\&C. Furthermore, the presented representative simulation results have verified the benefits of the proposed methods. As UAV-enabled ISAC remains largely unexplored, it is hoped that this paper will provide a useful initial guide and motivation for future research. 

\footnotesize  	
\bibliography{mybibfile}
\bibliographystyle{IEEEtran}

\normalsize
\vspace{10mm}

\section{Biographies}

{\bf{Kaitao Meng}} is currently a Post-Doctoral Researcher with the State Key Laboratory of Internet of Things for Smart City, University of Macau, Macau, China. His current research interests include integrated sensing and communication, multi-UAV collaboration, and intelligent reflecting surface.

\vspace{3mm}

{\bf{Qingqing Wu}} is currently an Associate Professor with the Department of Electronic Engineering, Shanghai Jiao Tong University, China. He was listed as the Clarivate ESI Highly Cited Researcher
in 2021 and 2022, the Most Influential Scholar Award in AI-2000 by Aminer in 2021, and World's Top 2\% Scientist by Stanford University in 2020 and 2021.

\vspace{3mm}

{\bf{Jie Xu}} is currently an Associate Professor with the School of Science and Engineering, The Chinese University of Hong Kong, Shenzhen, China. His research interests include wireless communications, wireless information and power transfer, UAV communications, edge computing and intelligence, and integrated sensing and communication (ISAC). He served or is serving as an Editor of IEEE Transactions on Wireless Communications, IEEE Transactions on Communications, and IEEE Wireless Communications Letters.

\vspace{3mm}

{\bf{Wen Chen}} is a tenured Professor with the Department of Electronic Engineering, Shanghai Jiao Tong University, China, where he is the director of Broadband Access Network Laboratory. He is a fellow of Chinese Institute of Electronics and the distinguished lecturers of IEEE Communications Society and IEEE Vehicular Technology Society. He is the Shanghai Chapter Chair of IEEE Vehicular Technology Society, Editors of IEEE Transactions on Wireless Communications, IEEE Transactions on Communications, IEEE Access and IEEE Open Journal of Vehicular Technology. %His research interests include multiple access, wireless AI and meta-surface communications. He has published more than 130 papers in IEEE journals and more than 120 papers in IEEE Conferences, with citations more than 9000 in google scholar.

\vspace{3mm}

{\bf{Zhiyong Feng}} is a professor at Beijing University of Posts and Telecommunications (BUPT), and the director of the Key Laboratory of the Universal Wireless Communications, Ministry of Education, P.R.China. Currently, she is serving as Associate Editors-in-Chief for China Communications. Her main research interests include wireless network architecture design and radio resource management in mobile networks, spectrum sensing and dynamic spectrum management in cognitive wireless networks, and integrated sensing and communications.

\vspace{3mm}

{\bf{Robert Schober}} is an Alexander von Humboldt Professor and the Chair for Digital Communication at Friedrich-Alexander University of Erlangen-Nuremberg (FAU), Germany,. His research interests fall into the broad areas of Communication Theory, Wireless and Molecular Communications, and Statistical Signal Processing.

\vspace{3mm}

{\bf{A. Lee Swindlehurst}} received the B.S. and M.S. degrees in Electrical Engineering from BYU, and the PhD degree in Electrical Engineering from Stanford. He was with the ECE Department at BYU from 1990-2007, then on leave from during 2006-07 working as VP of Research for ArrayComm LLC. Since 2007 he has been a Professor in the EECS Department at UC Irvine. During 2014-17 he was also a Hans Fischer Senior Fellow at the Technical University of Munich. He is an IEEE Fellow, and in 2016 he was elected as a Foreign Member of the Royal Swedish Academy of Engineering Sciences. 

\vspace{3mm}

\end{document}